\documentclass[prb,byrevtex,floatfix,twocolumn,superscriptaddress]{revtex4}
\usepackage{color}
\usepackage{amssymb,graphicx,afterpage}
\begin{document}
\title{\boldmath Malaria pigment crystals as magnetic micro-rotors: key for high-sensitivity diagnosis \unboldmath}
%
%
\author{A. Butykai}
\affiliation{Department of Physics, Budapest University of
Technology and Economics and Condensed Matter Research Group of the
Hungarian Academy of Sciences, H-1111 Budapest, Hungary}
\author{A. Orb\'an}
\affiliation{Department of Physics, Budapest University of
Technology and Economics and Condensed Matter Research Group of the
Hungarian Academy of Sciences, H-1111 Budapest, Hungary}
\author{V. Kocsis}
\affiliation{Department of Physics, Budapest University of
Technology and Economics and Condensed Matter Research Group of the
Hungarian Academy of Sciences, H-1111 Budapest, Hungary}
\author{D. Szaller}
\affiliation{Department of Physics, Budapest University of
Technology and Economics and Condensed Matter Research Group of the
Hungarian Academy of Sciences, H-1111 Budapest, Hungary}
\author{S. Bord\'acs}
\affiliation{Department of Physics, Budapest University of
Technology and Economics and Condensed Matter Research Group of the
Hungarian Academy of Sciences, H-1111 Budapest, Hungary}
\author{E. T\'atrai-Szekeres}
\affiliation{Department of Physics, Budapest University of
Technology and Economics and Condensed Matter Research Group of the
Hungarian Academy of Sciences, H-1111 Budapest, Hungary}
\author{L.F. Kiss}
\affiliation{Institute for Solid State Physics and Optics, Wigner
Research Centre for Physics, Hungarian Academy of Sciences, H-1525
Budapest, Hungary}
\author{A. B\'ota}
\affiliation{Department of Biological Nanochemistry, Institute of
Molecular Pharmacology, Research Center for Natural Sciences,
Hungarian Academy of Sciences, H-1025 Budapest, Hungary}
\author{B.G. V\'ertessy}
\affiliation{Institute of Enzymology, Research Center for Natural
Sciences, Hungarian Academy of Sciences, H-1113 Budapest, Hungary}
\affiliation{Department of Applied Biotechnology and Food Science,
Budapest University of Technology and Economics, H-1111 Budapest,
Hungary}
\author{T. Zelles}
\affiliation{Department of Oral Biology, Semmelweis University,
H-1089 Budapest, Hungary}
\author{I. K\'ezsm\'arki*}
\affiliation{Department of Physics, Budapest University of
Technology and Economics and Condensed Matter Research Group of the
Hungarian Academy of Sciences, H-1111 Budapest, Hungary}
\date{\today}
\begin{abstract}
The need to develop new methods for the high-sensitivity diagnosis
of malaria has initiated a global activity in medical and
interdisciplinary sciences. Most of the diverse variety of emerging
techniques are based on research-grade instruments, sophisticated
reagent-based assays or rely on expertise. Here, we suggest an
alternative optical methodology with an easy-to-use and
cost-effective instrumentation based on unique properties of malaria
pigment reported previously and determined quantitatively in the
present study. Malaria pigment, also called hemozoin, is an
insoluble microcrystalline form of heme. These crystallites show
remarkable magnetic and optical anisotropy distinctly from any other
components of blood. As a consequence, they can simultaneously act
as magnetically driven micro-rotors and spinning polarizers in
suspensions. These properties can gain importance not only in
malaria diagnosis and therapies, where hemozoin is considered as
drug target or immune modulator, but also in the magnetic
manipulation of cells and tissues on the microscopic scale.
\end{abstract}
\maketitle
%
%

In spite of the global efforts made for its elimination including
preventive strategies and drug therapies, malaria is still the
topmost vector-borne infectious disease with more than 200 million
clinical cases and around 1 million fatalities a year\cite{WHO2011}.
Increasing drug resistance of the parasites strongly acts against
the global malaria control, while climate change can even result in
the reintroduction of malaria mosquitos into post-endemic countries.
A significant improvement could be achieved via the development of
cheap diagnostic methods accurate even at the early stage of the
infection and via new drugs or vaccines efficient against the most
severe types of malaria parasites\cite{Greenwood2005,Yuan2011}.

Among diagnostic methods currently in practice the most reliable and
sensitive one is the microscopic observation of blood smears --able
to detect parasitemia associated with 5-10 parasites in 1\,$\mu$l
blood--, which is rather costly as requiring expertise and
high-powered microscopes. Though antigen-based detection of malaria
parasites offers a cheaper alternative and the corresponding rapid
diagnostic tests (RDT) are widely
used\cite{Moody2002,Bell2006,Wilson2012}, presently these techniques
have strong limitations. Perhaps the two major drawbacks are that i)
RDTs are not sensitive enough to detect early-stage infections, the
current sensitivity threshold being around 100\,parasites$/$$\mu$l
and ii) the tests are not quantitative enough to distinguish between
levels of infections (in endemic areas, most individuals will test
positive showing some degree of parasitemia but not allowing
identification of patients with active disease requiring urgent
treatment)\cite{Murray2009}. Additionally, false positive results
may arise due to the imperfect clearage of antigen proteins from the
body after successful treatments, while false negative results are
owing to their absence in certain Plasmodium
strains\cite{Gamboa2010}, as is the case for histidine-rich protein
II\cite{Rock1987}. Although among the molecular biology-based
methods, polymerase chain reaction (PCR) assays are sensitive enough
to detect 1 parasite\,$/$$\mu$l \cite{Andrade2010,Khairnar2009}, the
practical use of PCR assays on the field is limited due to
requirements of sophisticated technology and expertise.

In the last few years, the need to develop new diagnostic methods
has been driving extended research and a large arsenal of diagnostic
schemes has been proposed. Some of them are still based on selective
microscopic detection of infected blood cells such as magnetic
deposition microscopy\cite{Zimmerman2006}, third harmonic generation
imaging\cite{Belisle2008}, fluorescent study of cell microarray
chips\cite{Yatsushiro2010} and photoacoustic
flowmetry\cite{Samson2012}. There is an increasing number of methods
using malaria pigment as the target material for magnetic diagnosis
including magnetic purification
processes\cite{Zimmerman2006,Kim2010}, magneto-optical detection
using polarized light\cite{Newman2008}, electrochemical magneto
immunosensors\cite{Castilho2011} and magnetic field enriched surface
enhanced resonance Raman spectroscopy\cite{Yuen2012}.

Malaria pigment, also called hemozoin, is a byproduct of the disease
formed during the intraerythrocytic growth cycle of the
parasites\cite{Goldberg1990,Gluzmann1994}. Digestion of hemoglobin
by the malaria parasites results in the accumulation of monomeric
heme. As it is highly toxic to the parasites, they transform heme
into an insoluble crystallized form in which heme groups are
dimerized through iron–carboxylate links and the three dimensional
structure is stabilized via hydrogen bonds\cite{Slater1991}. This
process is accompanied by the change in the valency and the local
coordination of iron and leads to the transformation of low-spin
diamagnetic Fe$^{2+}$ ions contained in oxyhemoglobin into high-spin
(S=5/2) paramagnetic Fe$^{3+}$ ions in
hemozoin\cite{Bohle1998,Sienkiewicz2006,Walczak2005}. Besides
playing a key role in several diagnostic techniques, malaria
pigment, an ensemble of submicron-sized paramagnetic hemozoin
crystallites, is a main drug target and may also act as an immune
modulator\cite{Hanscheid2007,Jamarillo2009}.

\begin{figure}[h!]
\includegraphics[width=3.4in]{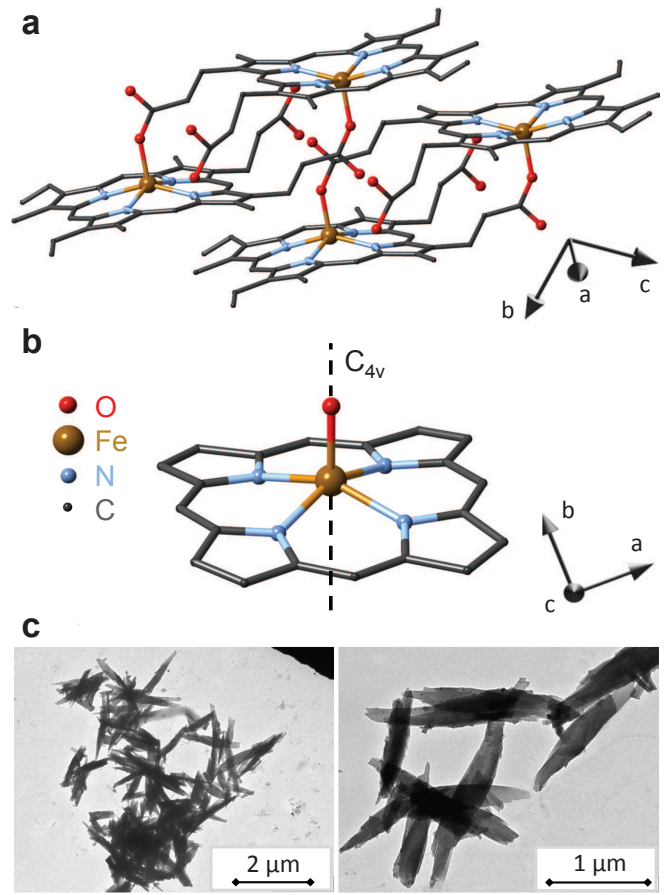}
\caption{\textbf{$\mid$ Structure and morphology of hemozoin
crystals.} (\textbf{a}) Triclinic structure of hemozoin with two
unit cells displayed using structural data from
Ref.~\onlinecite{Pagola2000}. The main crystallographic axes $a$,
$b$ and $c$ are also indicated. (\textbf{b}) The local symmetry of
five-fold coordinated iron in hemozoin nearly preserves a four-fold
rotation axis, $C_{4v}$. The angle spanned by this C$_{4v}$ axis
(hard axis of the magnetization) and the crystallographic $c$-axis
(fore-axis of the elongated crystals) is $\delta$$\approx$60$^o$,
where the $c$-axis points out of the plane of the figure.
(\textbf{c}) Transmission electron micrographs of typical hemozoin
crystallites dried from suspensions.} \label{fig:structure}
\end{figure}

Hemozoin has a low-symmetry triclinic crystal
structure\cite{Pagola2000} as shown in Fig.~\ref{fig:structure}a.
Although the morphology of the crystallites shows variations
depending on the parasite species\cite{Noland2003,Oliveiraa2005},
they typically have an elongated rod-like shape with a length
ranging from 300\,nm to 1\,$\mu$m. Besides the natural formation of
hemozoin inside the parasites, various methods have been established
for its chemical synthesis\cite{Slater1991,Bohle1993}. Though the
artificially grown version is usually called $\beta$-hematin to be
distinguished from hemozoin, they are demonstrated to share
identical chemical composition, crystal
structure\cite{Slater1991,Pagola2000},
optical\cite{Slater1991,Frosch2007} and magnetic
properties\cite{Bohle1998,Sienkiewicz2006}. Hereafter, we will refer
to both natural and synthetic versions of malaria pigment as
hemozoin. Hemozoin crystals used in the present study were prepared
from hemin by an aqueous acid-catalyzed
reaction\cite{Jamarillo2009}. Transmission electron micrograph (TEM)
images of typical crystallites are shown in
Fig.~\ref{fig:structure}c. Similarly to the naturally grown ones,
they are elongated and characterized by a size distribution of
$\sim$700$\pm$200\,nm.

Here, by extending the work of Newman and coworkers
\cite{Newman2008} we report a new path for the magnetic detection of
malaria, which, in contrast to most of the aforementioned recently
emerging techniques, may be realized as a cheap and compact
diagnostic tool without the application of research-grade
instruments. Unique magnetic and optical properties of malaria
pigment crystals as well as their highly controllable dynamics in
fluids, which all play key roles in the principle of detection, are
systematically investigated. The threshold of our device detecting
hemozoin content, as in the present trial state, is 15\,picogram of
hemozoin in 1\,$\mu$l blood equivalent to a level of parasitemia
$\lesssim$30\,parasites$/$$\mu$l \cite{Newman2008}, which needs to
be confirmed via extended clinical trials.  While this detection
limit already shows improvement over that of RDTs, it is further
decreased by about more than one order of magnitude for hemozoin
detection in blood plasma or serum corresponding to a parasitemia
less than 1\,parasite$/$$\mu$l.

\begin{figure*}[t!]
\includegraphics[width=7.05in]{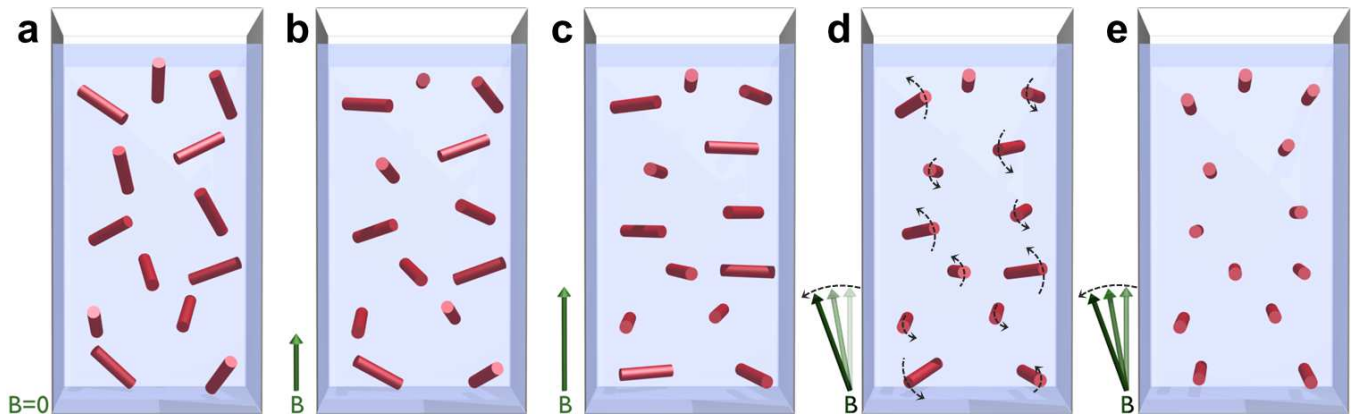}
\caption{\textbf{$\mid$ Magnetic orientation and dynamics of
paramagnetic hemozoin crystals with anisotropic easy-plane
character.} In these schematic drawings, the cylinders represent the
suspended hemozoin crystals. The axes of the cylinders correspond to
the magnetic hard axes of the crystals and not related to their
fore-axes. (\textbf{a}) Without external magnetic field the crystals
in the suspension are randomly oriented. (\textbf{b}) With the
application of a magnetic field, the hard axes of the crystals begin
to align perpendicular to the magnetic field vector $\textbf{B}$,
though this orientation is hindered by the thermal fluctuations.
(\textbf{c}) In the high-field limit this two-dimensional alignment
is completed, with the hard axis of each crystal lying within the
plane normal to the field. (\textbf{d}) In slowly rotating fields
the crystallites behave as magnetically driven micro-rotors.
(\textbf{e}) Due to the viscosity of the fluid, at high rotation
frequencies their hard axes tend to align parallel to the rotation
axis and consequently they stop spinning. Only in this case a full
three dimensional alignment of the hard axes is achieved.}
\label{fig:orientation}
\end{figure*}

\section*{Results}
The diagnostic methodology presented in this paper relies to a large
extent on magnetic properties which are highly specific to malaria
pigment crystals and unique in human body. First we review the
fundamental characteristics of the crystallites determined by
magnetization and magneto-optical measurements. The following part
describes the dynamics of their magnetically driven rotation in
fluids with different viscosity such as hemolyzed blood, water,
acetone, etc.

\textbf{Magnetic anisotropy of hemozoin.} The low crystal symmetry
would generally imply that hemozoin is a highly anisotropic
paramagnet with different magnetic susceptibility values along each
of the three main crystallographic axes. However, Fe$^{3+}$ ions
located in the center of porphyrin rings experience higher local
symmetry since the four-fold rotational axis perpendicular to the
plane of the porphyrin unit is nearly preserved as shown in
Figs.~\ref{fig:structure}a-b. Therefore, we expect that the magnetic
properties of malaria pigment, which are mainly determined by
Fe$^{3+}$ ions, reflect this axial (C$_{4v}$) symmetry and hemozoin
behaves either as an easy-axis or as an easy-plane paramagnet.

Based on a multi-frequency high-field electron paramagnetic
resonance (EPR) study on powder samples, Sienkiewitz and
co-workers\cite{Sienkiewicz2006} suggested that the behaviour of the
S=5/2 spins of Fe$^{3+}$ ions in hemozoin can be described by the
following Hamiltonian, $H= D \left( S_{z}^2-\frac{S(S+1)}{3}\right)
+ E(S_{x}^2-S_{y}^2) + \mu_B g \mathbf{BS}$, where $D$ in the first
term is the zero-field splitting associated with an axial
anisotropy, while lowering of the C$_{4v}$ symmetry introduced via
$E$ is negligible as $|E/D|\leq0.035$. They also found that the
Zeemann-splitting induced by an external magnetic field,
$\mathbf{B}$, is characterized by a nearly isotropic g-factor,
g$\approx$2. The dominance of the $D$ term together with its
positive sign found in this low-temperature EPR study -- in
agreement with the results obtained by M\"ossbauer
spectroscopy\cite{Bohle1998} -- hint toward the fact that hemozoin
can behave as an easy-plane paramagnet over an extended temperature
region. According to the local symmetry generated by the ligand
field at iron sites, we assign the easy plane of the magnetization
with the plane of the porphyrin rings, hence, the hard direction
labeled as $z$-axis in the spin Hamiltonian above coincides with the
four-fold rotational axis.

When such easy-plane crystallites suspended in a liquid are exposed
to an external magnetic field, they tend to co-align with the field
direction to gain magnetic energy (see
Figs.~\ref{fig:orientation}a-c). Assuming a linear field dependence
of the magnetization, which is experimentally confirmed at room
temperature, the magnetic anisotropy energy is
$U$$=$$-\frac{1}{2}\frac{B^2}{\mu_0}cos^2(\theta)(\chi_{zz}-\chi_{xx})V$.
Here, $\chi_{zz}$ and $\chi_{xx}$ stand for the linear magnetic
susceptibility of a crystal along the hard axis and within the easy
plane, respectively, $\theta$ is the angle between the direction of
the field and the hard axis of a crystal and $V$ is its volume.
Magnetization densities for fields applied within the easy plane and
along the hard axis of the crystal are given by
$M_x=\chi_{xx}B/\mu_0$ and $M_z=\chi_{zz}B/\mu_0$, respectively.
(For details see Methods section.) Since thermal fluctuations try to
restore the random orientation, the angular distribution of the
crystals over the suspension depends on the relative strength of the
magnetic anisotropy energy and the energy scale of thermal
fluctuations, $k_BT$, according to
$f(\theta)=\frac{e^{-U/k_BT}}{2\pi
\int_{0}^{\pi}e^{-U/k_BT}\sin{\theta}d\theta}$.

In order to directly determine the strength of the magnetic
anisotropy we car
ried out field- and temperature-dependent
magnetization measurements on powder samples of randomly oriented
crystals as well as on crystals suspended in a mixture of 70\% water
and 30\% glycerol. In the second case, the measurements were
performed both after zero-field cooling for maintaining the random
orientation of the crystals in the suspension and after a field
cooling process used to magnetically align the crystals and fix them
by freezing the mixture. (Fixation occurs below the freezing point
of the mixture, $T_{fr}$$\approx$230\,K -- see Supplementary
Information.) For suspensions with hemozoin content less than
10\,$\mu$g/$\mu$l, the magnetic field of B$=$5\,T used for field
cooling from room temperature down to T$=$2\,K was found safely
large to achieve high degree of orientation within the suspensions
as schematically illustrated in Fig.~\ref{fig:orientation}c. This
way we could measure the magnetization specific to the case when the
magnetic field lies within the easy plane and the hard axis of each
crystal is aligned perpendicular to the field.

\begin{figure}[h!]
\includegraphics[width=3in]{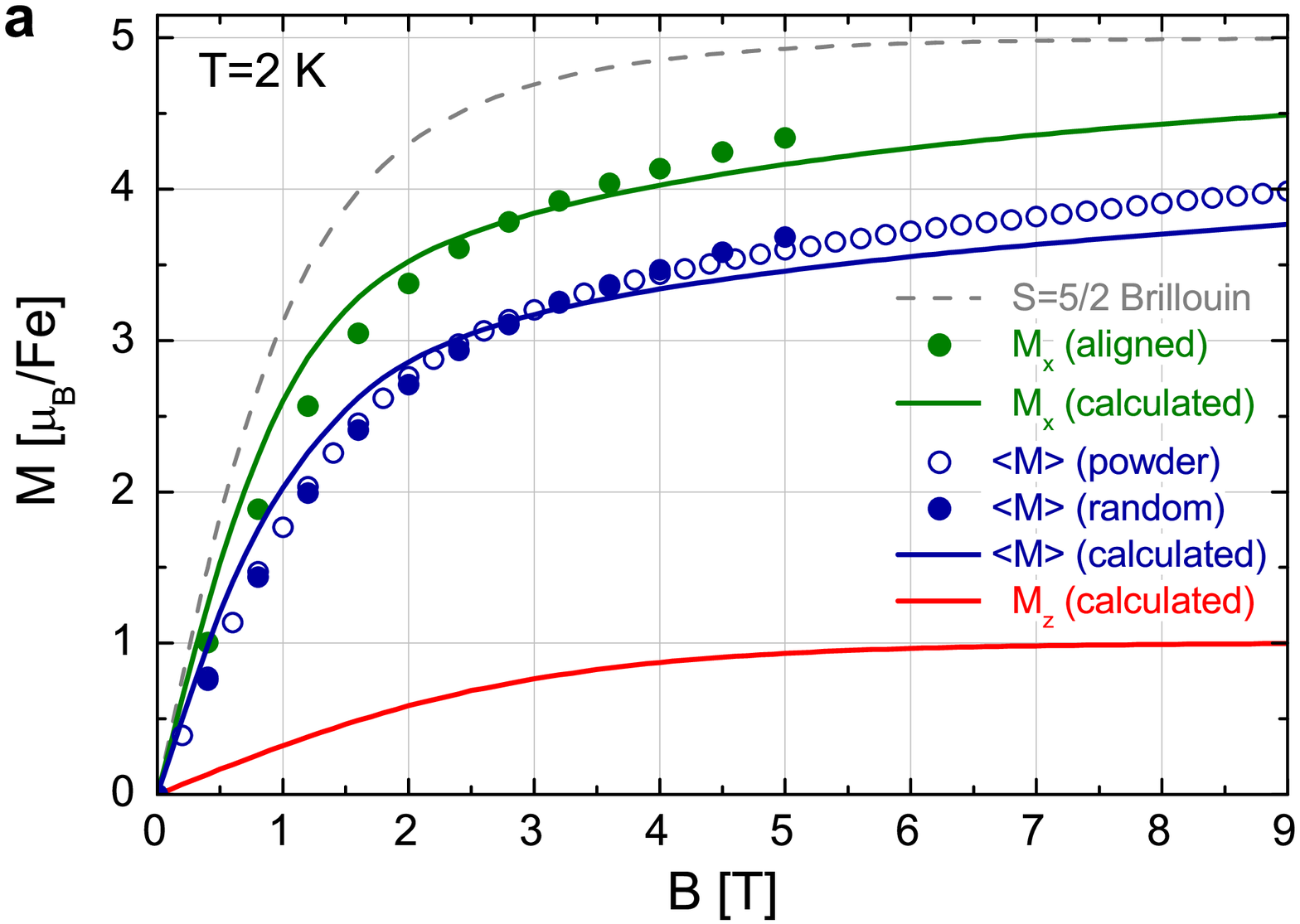}\\
\vspace{0.1in}
\includegraphics[width=3in]{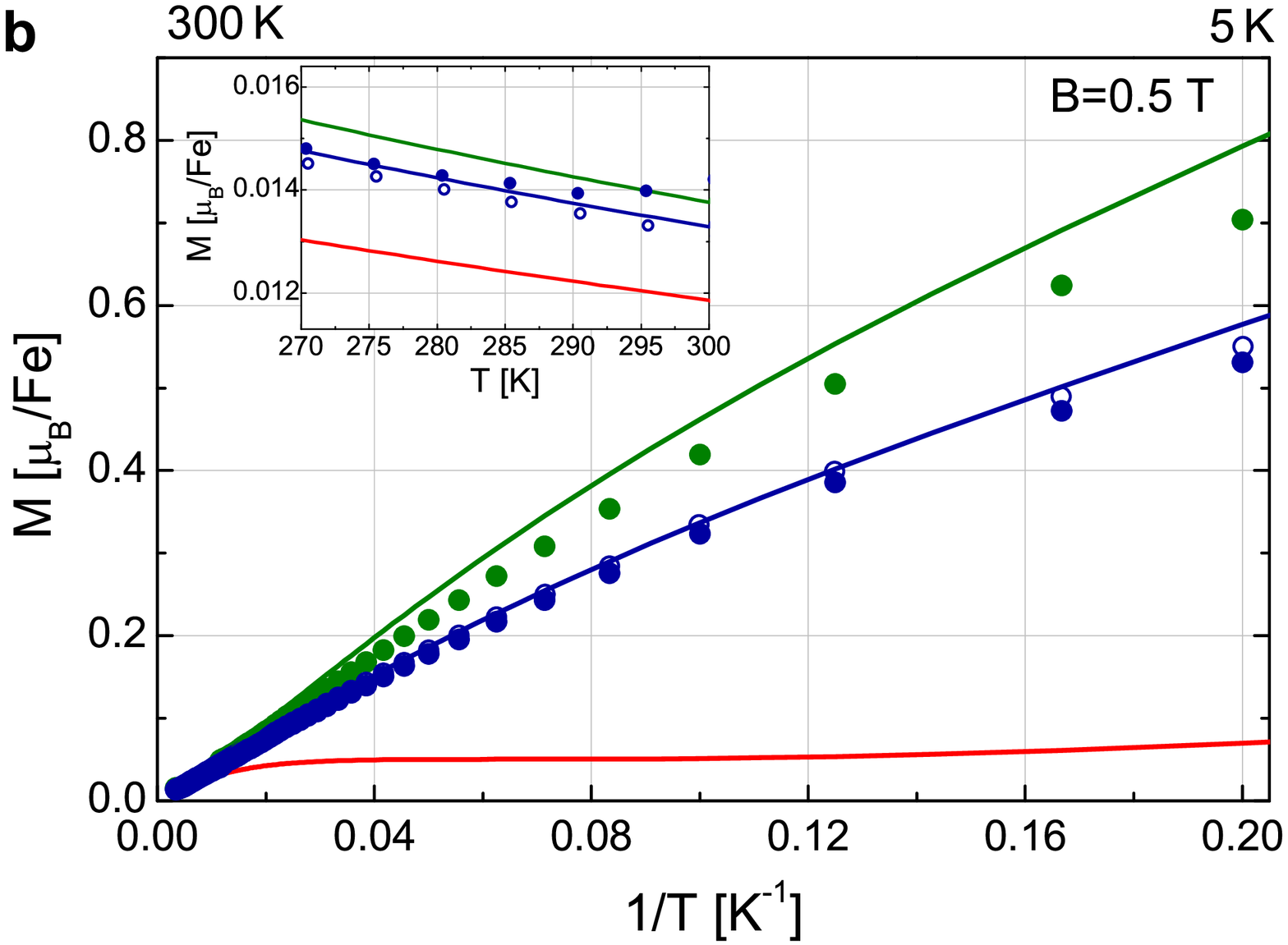}
\caption{\textbf{$\mid$ Magnetization anisotropy of malaria pigment
crystals.} (\textbf{a}) Field dependence of the magnetization
measured at T$=$2\,K for a powder sample, randomly oriented
(zero-field cooled suspension) crystals and magnetically aligned
(field cooled suspension) crystals are shown by blue open circles,
blue dots and green dots, respectively. As expected, the former two
are essentially identical. Magnetization curves calculated for
fields lying within the easy plane and pointing along the hard axis
of a crystal are also plotted with green and red lines,
respectively. The angular average of the magnetization corresponding
to the random orientation of the crystals is also displayed with
blue line. (For details of the calculation see the Methods section.)
Magnetization values are given for a single iron site in
Bohr-magneton units. To emphasize the anisotropic character of
hemozoin, Brillouin's function describing the magnetization of an
isotropic S=5/2 spin is also shown (dashed grey line). (\textbf{b})
Low-field magnetization of hemozoin as a function of the inverse
temperature measured in B=0.5\,T. The position of 300\,K and 5\,K
are indicated on the upper scale. The inset shows the data on a
linear temperature scale around 300\,K. Symbols and lines indicate
respectively the same measured and calculated quantities as in panel
(\textbf{a}).} \label{fig:magnetization}
\end{figure}

Here, we note that the magnetic hard axis is roughly parallel to the
[131] crystallographic axis, while the crystallites are elongated
along the [001] direction\cite{Buller2002,Kapishnikov2012}.
Therefore, the two dimensional co-alignment of the hard axes of the
crystals (shown in Fig.~\ref{fig:orientation}c) leaves a large
freedom for the orientation of their fore-axis since the crystals
can rotate around both their hard axes and the direction of the
external magnetic field without any change in their magnetic energy.
This is in agreement with our TEM observations (see Supplementary
Information), which showed that magnetic alignment is not
straightforwardly manifested in the orientation of the shape of the
crystals, in contrast to former assumptions \cite{Newman2008}.

The field dependence of the magnetization at T$=$2\,K and the
temperature-dependent magnetization measured in B$=$0.5\,T with
increasing temperature are plotted in Fig.~\ref{fig:magnetization}a
and \ref{fig:magnetization}b, respectively, for powder samples,
zero-field cooled and field cooled suspensions. The difference
between the magnetization of the randomly oriented and the aligned
samples increases with decreasing temperature and clearly shows the
anisotropic nature of hemozoin.

To determine the magnetization also for fields pointing along the
hard axis of a given crystal from these data, we calculated the
magnetization, both as a function of field and temperature, using
the spin Hamiltonian with axial anisotropy and tuned the value of
$D$ to obtain the best fitting with the measured curves. We found
that the magnetization of a crystal strongly depends on the angle
$\theta$ spanned by its hard axis and the direction of the external
field. Besides the magnetization density values corresponding to the
easy plane ($M_x$) and the hard axis ($M_z$), the magnetization of a
sample containing randomly oriented crystals, $\langle M\rangle$,
was also evaluated by averaging over $\theta$. All the experimental
data shown in Fig.~\ref{fig:magnetization} and additional data
presented in the Supplementary Information are well reproduced by a
common value of the single fitting parameter, $D$=13.4\,K, which is
consistent with the value reported in former
EPR\cite{Sienkiewicz2006} and M\"ossbauer\cite{Bohle1998}
spectroscopic studies.

For the anisotropy of the low-field magnetization (or alternatively
the linear susceptibility) we obtained a value as large as
$M_x/M_z=9.6\pm0.2$ at T$=$2\,K. Though this ratio is gradually
reduced when the energy scale of the thermal fluctuations becomes
comparable and larger than the zero field splitting, i.e. for
$k_BT>D$, it is still considerable at T$=$300\,K with
$M_x/M_z\approx1.16\pm0.03$. The magnitude of the anisotropy at room
temperature implies that partial (two dimensional) and full (three
dimensional) magnetic alignment of the crystals respectively
obtained by static and rotating fields can be achieved by magnetic
fields of $\lesssim$1\,T as proposed in Fig.~\ref{fig:orientation}.

\textbf{Magnetically induced linear dichrosim of malaria pigment.}
Similarly to the magnetic anisotropy, the planar stacking of
$Fe^{3+}$-protoporphyrin-IX units in hemozoin\cite{Pagola2000}
together with the axial symmetry of iron sites are indicative of
anisotropic optical properties for a single crystal (see
Figs.~\ref{fig:structure}a-b). More specifically, optical
excitations in the absorption spectrum of hemozoin over the
near-infrared and the visible regions can be assigned to transitions
mainly involving $\pi$ and $\pi^*$ orbitals of the porphyrin and $d$
orbitals of the central $Fe^{3+}$ ion\cite{Eaton1978,Wood2004}.
These assignments also support the local C$_{4v}$ symmetry of iron
in hemozoin similarly to the case of hemin and
deoxyhemoglobin\cite{Eaton1978}. Since the same symmetry dictates
the magnetic and optical anisotropy of hemozoin on the microscopic
level, alignment of the crystallites by external field is expected
to simultaneously generate macroscopic magnetic and optical
anisotropy in their suspensions.

\begin{figure}[t!]
\includegraphics[width=3in]{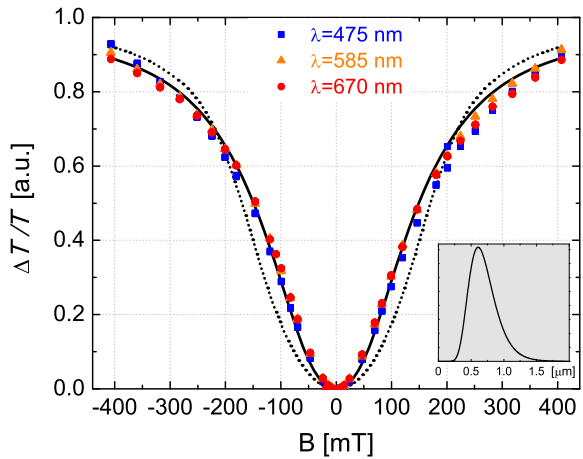}
\caption{\textbf{$\mid$ Magnetically induced linear dichroism in
hemozoin suspensions.} Magnetic field dependence of linear dichroism
measured on a room-temperature aqueous suspension of hemozoin at
wavelengths $\lambda=$475\,nm (blue squares), 585\,nm (orange
triangles) and 670\,nm (red dots). Data corresponding to different
wavelengths were normalized to a common scale, which resulted in a
universal field dependence reproduced well by the theory. Assuming
an average-sized crystal, the fitting (dotted line) yields
$M_x/M_z$$=$1.11 for the magnetization anisotropy, which corresponds
to $M_x-M_z$$=$$0.013$\,$\mu_B$/Fe in a magnetic field of 5\,T. The
quality of the fit can be further improved (solid line) by assuming
the distribution of crystal size shown in the caption. For details
see the main text.} \label{fig:MLD}
\end{figure}

This invokes a diagnostic tool based on magneto-optical phenomena
such as magnetically induced linear birefringence/dichroism or
polarization dependent light scattering. Though all of these three
effects can be relevant we will refer to them as magnetically
induced linear dichrosim (MLD). Recently, Newman and coworkers have
reported a magneto-optical methodology capable of a sensitive
diagnosis of malaria\cite{Newman2008,Mens2010}. They found a
specific field dependence of MLD\cite{Newman2008_2}. In order to
probe the microscopic properties of hemozoin we revisited this
phenomenon and studied its wavelength dependence.

We investigated the magnetic field dependence of the linear
dichroism on aqueous suspensions of hemozoin at multiple wavelengths
(e.g. $\lambda$=475\,nm, 585\,nm and 670\,nm) by measuring the
transmitted intensity in Voigt configuration for light polarizations
parallel and perpendicular to the applied field using a polarization
modulation technique. (See Methods section for details.) The
experimental curves obtained at different wavelengths follow a
universal field dependence when normalized to a common scale as
shown in Fig.~\ref{fig:MLD}. After the quadratic increase of MLD at
low fields, the signal tends to saturate with an inflection at an
intermediate field B$_0$$\approx$0.1\,T.

The mechanism behind MLD in hemozoin suspensions is outlined
schematically in Figs.~\ref{fig:orientation}a-c. In a dilute
suspension the crystals are oriented randomly, resulting in an
optically isotropic media since polarization effects from individual
crystals average to zero. As already discussed, in external magnetic
fields the crystals align in a manner that the field would
preferably lie within their easy planes. The ordering is opposed by
thermal fluctuations and this competition determines the specific
field dependence of MLD. The linear dichroism of a single hemozoin
crystal is characterized by the difference of its transmission
coefficients $T_x$ and $T_z$ corresponding to light polarizations
parallel and perpendicular to the porphyrin planes, respectively.
(These directions were respectively called easy plane and hard axis
in the former magnetic terminology.) In external magnetic field the
contributions from individual crystals produce a macroscopic
transmission anisotropy between polarizations parallel and
perpendicular to the field direction
\begin{equation}
\frac{\Delta \mathcal{T}}{\mathcal{T}}=
c\cdot\frac{T_x-T_z}{T_x+T_z}\cdot\int_{0}^{\pi} \pi
f(\theta)(3\cos^{2}\theta-1)\sin{\theta} d\theta, \label{LD}
\end{equation}
where $\Delta \mathcal{T}$ and $\mathcal{T}$ is the difference and
the average of the transmitted intensities for the two
polarizations, respectively. The factor $c$ expresses linear scaling
with the concentration. The wavelength dependence emerges in the
second term through the transmission anisotropy of individual
crystals, while the integral captures the field dependence,
hereafter referred to as $\Phi(B)$, describing the degree of the
magnetic alignment \cite{Langevin1910}.

The universal field dependence shown in Fig.~\ref{fig:MLD} was
fitted by numerically evaluating $\Phi(B)$ for different values of
the magnetic anisotropy using a typical crystal size of
V=200$\times$200$\times$700\,nm$^3$. We found the best fitting with
$M_x/M_z=1.11\pm0.04$ in good agreement with the value obtained from
the magnetization study on oriented samples at room temperature. As
the difference of the experimental and the fitted curves are likely
due to the size-distribution of the crystals, we refined the fit
assuming lognormal distribution of the crystal size. To reduce the
number of free parameters, we fixed the average length of the
crystals and the magnetization anisotropy to the values previously
obtained by the single-size-fit, i.e. L$=$700\,nm and $M_x/M_z=1.1$.
Then, we obtained 220\,nm for the standard deviation of the length
and 8:2 for the aspect ratio of the crystals. These are both
realistic in the light of scanning electron micrographs (aspect
ratio of 7:2 was considered in the single-size-fit) and further
improved the quality of the fit.

\textbf{Spectral features of the MLD effect in hemozoin.} To gain
more insight into the microscopic optical properties of hemozoin and
trace the spectral range optimal for diagnosis, we studied MLD
effect from the ultraviolet to the near-infrared region
($\lambda$$=$300-1300\,nm) in B=0.3\,T. We also investigated the
influence of different suspension media, including normal saline,
blood plasma and blood, on the detectability of hemozoin.

\begin{figure}[t!]
\includegraphics[width=3.2in]{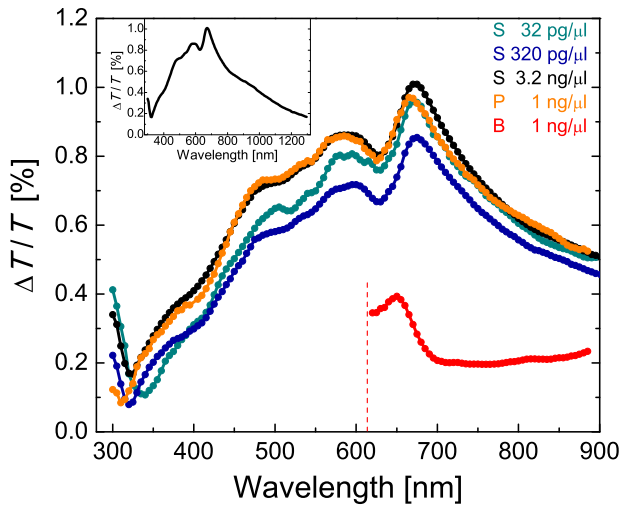}
\caption{\textbf{$\mid$ Wavelength dependence of the MLD effect in
hemozoin suspensions.} MLD spectra for room-temperature hemozin
suspensions in normal saline (S), blood plasma (P) and full blood
(B) normalized to a concentration of 1\,ng/$\mu$l. The inset shows
the MLD effect over a broader spectral range in normal saline.}
\label{fig:spectra}
\end{figure}

As shown in Fig.~\ref{fig:spectra} the MLD spectra of hemozoin
suspensions in saline and blood plasma are essentially identical and
exhibit characteristic peaks distributed mostly over the visible
range, which may serve as optical fingerprints of malaria pigment.
These peaks are likely dominated by the linear dichroism of
absorption bands observed over the same
range\cite{Bohle1993,Wood2004}. The magnitude of MLD spectra shows
linear dependence on the hemozoin content over a wide range of
concentrations reassuring the feasibility of a quantitative
diagnosis. The effect is the largest at $\lambda$$\approx$670\,nm,
where the macroscopic transmission anisotropy reaches $\Delta
\mathcal{T}/\mathcal{T}$$=$1\% for the suspension with 1\,ng/$\mu$l
hemozoin content corresponding to an optical path of d$=$10\,mm. If
we assume that only the absorption of the crystals contribute to
MLD, i.e. polarization dependent light scattering can be neglected,
this value corresponds to a robust transmission anisotropy
$\frac{T_x-T_z}{T_x+T_z}$$\approx$40\% for a typical crystal and a
large difference in its absorption coefficients $\Delta
\alpha$$=$$\alpha_z-\alpha_x$$\approx$1.5$\cdot10^4$\,cm$^{-1}$.

The MLD effect decreases in blood by a factor of $\sim$3. The
overall sensitivity is further reduced owing to strong absorption
and light scattering by the blood components, mainly by red blood
cells. For wavelengths shorter than $\sim$620\,nm the transmitted
intensity drastically drops allowing no further observation of the
MLD signal. To approach the sensitivity level achieved for blood
plasma by visible light, we used hemolyzed blood in following
studies, which helps to strongly reduce light scattering. Please
note that the hemolysis of actually infected blood samples is also
favourable for the diagnosis, since the hemozoin portion still
contained within the erythrocytes can be released into the blood
plasma, hence becoming effectively detectable this way. We also work
on the development of simple techniques for the separation of
hemoglobin from hemolyzed blood, while keeping hemozoin within the
plasma, and on the selective filtering of hemozoin \cite{Butykai}.

\textbf{Malaria pigment crystals as magnetic micro-rotors.} For a
sensitive detection of weak polarization effects generated by low
amounts of hemozoin in infected blood, it is inevitable to use
polarization modulation as was already proposed by Newman and
coworkers\cite{Newman2008} and also applied in our magneto-optical
experiments. However, besides polarization modulation of the probing
light, -- in the special case of hemozoin crystals -- also magnetic
modulation of light polarization is conceivable.

\begin{figure}[t!]
\includegraphics[width=3in]{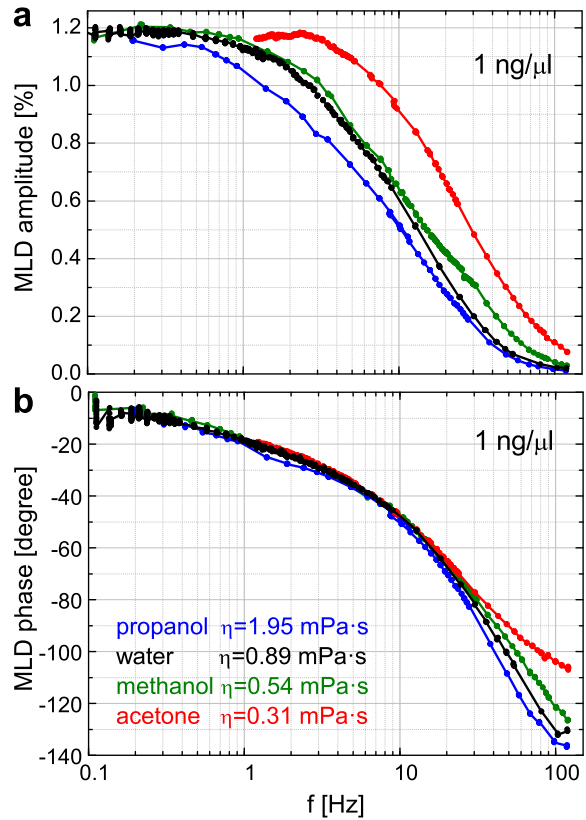}
\caption{\textbf{$\mid$ Magnetically driven dynamics of hemozoin
crystals in various suspension media at room temperature.}
(\textbf{a})/(\textbf{b}) Semi-logarithmic plot of MLD
amplitude/phase versus the frequency of the field rotation, $f$, for
hemozoin suspended in propanol, water, methanol and acetone with
1\,ng/$\mu$l concentration. Results in hemolyzed blood are
essentially identical with those obtained in water and not shown
here. Viscosity ($\eta$) for the different media are also
indicated.} \label{fig:freq}
\end{figure}

The central idea is that the magnetically aligned crystallites
follow the direction of a rotating magnetic field, i.e. they behave
as magnetically driven micro-rotors in a suspension. Moreover, the
dichroic planes of the crystals rotate in a synchronous manner, thus
the suspension acts as a spinning polarizer modulating the intensity
and the polarization of the transmitted light beam. The application
of a polarizing beam splitter after the sample and the differential
detection of the two orthogonally polarized beams by a balanced
photodiode-bridge provide an efficient scheme for the reduction of
intensity noise, meaning that an ordinary laser diode is sufficient
as light source. From the differential signal the a.c. component
corresponding to the second harmonic of the rotation frequency is
selectively detected, which originates solely from MLD caused by the
rotation of the dichroic crystals and it is not affected by
parasitic intensity noise coming from optical, mechanical and
thermal instability of the device.

A special arrangement of permanent magnets in a ring-shaped
structure surrounding the sample, called
Halbach-cylinder\cite{Halbach1980}, is used to generate a uniform
magnetic field of B=1\,T at the sample position. The ring is rotated
by a  d.c. electric motor with a frequency adjustable over the range
of f=0.1-130\,Hz, resulting in a field which rotates within the
plane perpendicular to the light path. We found that the efficient
co-alignment of the crystals using the strong and nearly homogeneous
magnetic field of the Halbach-cylinder rotated with fairly large
frequencies plays crucial role in the sensitivity of our method and
leads significant improvements over previous magneto-optical
detection schemes \cite{Newman2008,Mens2010}. (For further details
on the device see Methods section).

Applying this new methodology, we carried out a phase-sensitive
detection of MLD on hemozoin suspensions. Besides the amplitude of
the signal, its time delay relative to the rotating field has also
been recorded. To understand the dynamics of the crystals, we
measured the frequency dependence of MLD over f=0.1-130\,Hz in
solvents with different viscosity (see Fig.~\ref{fig:freq}). The
amplitude of MLD remains constant at low frequencies, then decays
drastically towards higher frequencies. The phase shift of MLD grows
gradually with increasing frequency with a viscosity dependence
exposed more in the high-frequency region.

These results can be understood via the basic features of the highly
complex crystal dynamics as schematically shown in
Figs.~\ref{fig:orientation}c-e. In a strong static field, as in
Fig.~\ref{fig:orientation}c, the hard axes of the crystallites are
distributed uniformly in the plane perpendicular to the field
direction. Upon the rotation of this field, the resulting torque
forces the hard axes of the crystals to precess around the rotation
axis and follow the field. Due to the viscosity of the fluid the
system behaves as an ensemble of damped rotators. Consequently,
towards higher frequencies the crystals experience an increasing
angular delay relative to the field and their hard axes tend to
align parallel to the rotation axis. This manifests in the finite
phase and the decreasing amplitude of MLD, respectively. When this
alignment is completed, no magnetic torque acts on the crystals as
the magnetic field rotates within their easy planes, hence they stop
moving. The analysis of rotational dynamics for easy-axis and
easy-plane magnetic particles has recently been subject to extensive
theoretical and experimental research
\cite{Tierno_2009,Dhar_2007,Erb_2012_Soft}. The dynamics described
here is specific to crystals with easy-plane magnetic anisotropy as
was also reported for other easy-plane paramagnetic particles
\cite{Dhar_2007,Erb_2012_Soft,R. M. Erb} and fundamentally differs
from the motion of easy-axis crystallites. In the present easy-plane
situation, MLD signal is suppressed towards high frequencies because
optically isotropic planes of the dynamically co-aligned crystals
are exposed to the light and consequently no dichroism can emerge.

\begin{figure*}[t!]
\includegraphics[width=5in]{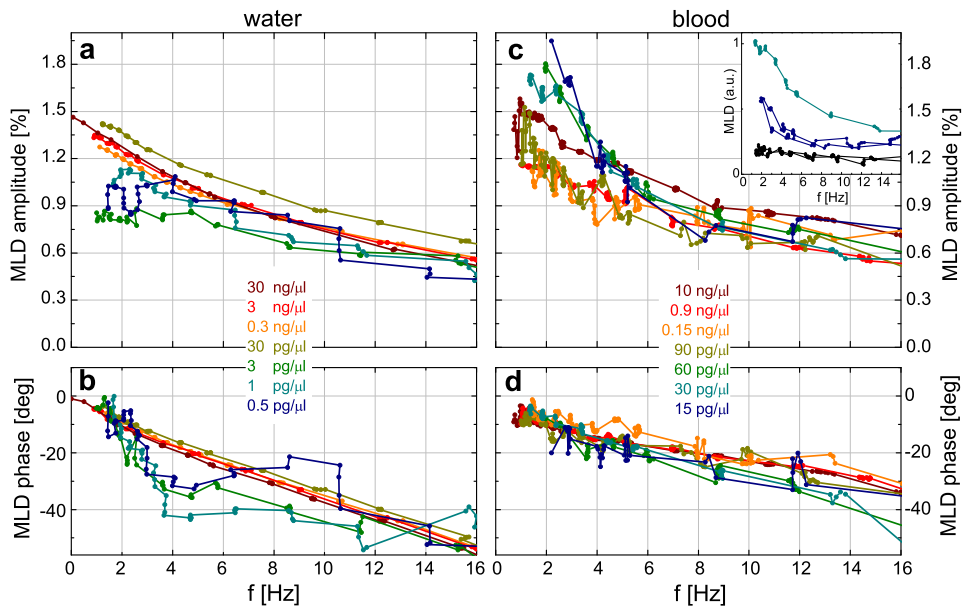}
\caption{\textbf{$\mid$ Sensitivity of the optical diagnostic method
based on the magnetic rotation of hemozoin crystals in water and
blood.} (\textbf{a})/(\textbf{b}) MLD amplitude/phase for hemozoin
in water over a limited frequency range optimal in sense of signal
to noise ratio. (\textbf{c})/(\textbf{d}) MLD amplitude/phase for
hemozoin in blood over the same frequency range. The concentration
of hemozoin varies over five and three orders of magnitude in water
and blood, respectively. The amplitude of the MLD signal is
normalized to 1\,ng/$\mu$l hemozoin content. Concentrations of blood
samples refer to the hemozoin contents in full blood and not in
hemolyzed blood. The concentration levels of 0.5\,pg/$\mu$l and
15\,pg/$\mu$l are still readily detectable in water and blood,
respectively. Inset in panel (\textbf{c}) shows the reproducibility
for the baseline (black curves) and the lowest-concentration data
(with color coding used in the main panel).} \label{fig:sens}
\end{figure*}

We tested the concentration threshold of hemozoin detection
achievable by this method. Frequency dependence of MLD was measured
for aqueous suspensions of hemozoin prepared over six orders of
magnitude in concentration, namely from $30$\,ng/$\mu$l to
$0.5$\,pg/$\mu$l. MLD curves displayed in Figs.~\ref{fig:sens}a-b
show that the lowest concentration corresponding to a parasitemia
less than 1\,parasite/$\mu$l is still readily detectable. More
relevant to diagnosis, our current detection limit for hemozoin in
blood is $c$$=$$15$\,pg/$\mu$l as demonstrated in
Figs.~\ref{fig:sens}c-d. This exceeds the performance of RDTs and
approaches the detection limit achievable by microscopic observation
of infected blood. In order to sufficiently reduce the strong light
scattering of red blood cells over the visible range, the
experiments were carried out in hemolyzed blood (see Methods
section). Please note, that the concentration values in
Figs.~\ref{fig:sens}c-d correspond to hemozoin contents in full
blood and not in hemolyzed blood obtained by 20-fold dilution with
distilled water. Thus, we can conclude that the sensitivity of the
detection in hemolyzed blood is close to that in water. The
precision of the hemozoin content for the series of blood samples
was checked by the parallel measurement of MLD signal on water with
the same hemozoin concentrations. At the moment the threshold of our
detection is not limited by the signal-to-noise ratio of MLD but by
a mainly frequency independent baseline superimposed on the signal.
This weak contribution to the second-harmonic signal may come e.g.
from the Voigt effect of the medium. The reproducibility of the
lowest-concentration data together with the baseline measured for
blood sample containing no hemozoin are displayed in the inset of
Figs.~\ref{fig:sens}c.

\section*{Discussion}
By combining magnetization measurements, broad-band magneto-optical
spectroscopy and electron transmission microscopy we determined
quantitatively the magnetic and optical anisotropy characteristic to
submicron-sized single crystals of hemozoin. Based on these results
we refined previous models describing the magnetic alignment of the
crystals in suspensions \cite{Newman2008,Mens2010}. These
fundamental properties offer a unique path for the magnetic
manipulation and optical detection of these crystallites and are
also relevant to the development of new drugs blocking hemozoin
production. Dielectric anisotropy may also play role in the
co-alignment of hemozoin crystallites during their nucleation within
the digestive vacuoles of the parasites\cite{Kapishnikov2012}.

We studied the dynamics of the crystals during their magnetically
driven rotation in suspensions with different viscosity including
hemolyzed blood. Based on the fact that the synchronous motion of
such micro-rotors induces intensity modulation of the transmitted
light via polarization effects, we assembled a device for the
detection of malaria pigment in blood. The device provides a more
sensitive way of diagnosis RDTs and approaches the sensitivity
achievable by microscopic observation of infected red blood cells,
which is the most effective diagnosis in practice to date. We found
that the sensitivity of the hemozoin detection in blood plasma is
even higher and we are aiming at the improvement of the detection
threshold by filtration techniques.

We expect no major reduction of sensitivity for infected blood
samples compared to the threshold reported here for synthetic
hemozoin in blood for the following reasons. First, hemolysis of
infected blood helps to release the portion of hemozoin still
contained within the erythrocytes into the blood plasma.
Furthermore, our preliminary data indicate that the magnitude of MLD
signal -- observed in large fields (B$\gtrsim$1\,T) rotating with
low frequencies -- shows only moderate changes with crystal size and
morphology in agreement with previous results \cite{Newman2008}. On
the other hand, the frequency dependence of both the amplitude and
phase of MLD may provide information specific to the type of the
parasites as increasing crystal size results in a shift of the decay
frequency towards higher values. Nevertheless, both the detection
threshold of our device and its specificity to different parasites
need to be proved via extensive clinical tests.

The use of hemozoin as the marker compound for detecting Plasmodium
infections has clear advantages over the currently used RDTs. The
production of hemozoin is a defense reaction on behalf of the
parasite that transforms a host protein into a magnetically
detectable compound with physico-chemical properties invariable upon
genetic variations of the parasites. Therefore, efficiency of our
method is not affected by high rates of their genetic variation. In
contrast, performance of the antigen-based diagnostics relies on the
antigen-antibody reaction which may be perturbed upon mutations in
the antigen. From this respect, it is important to emphasize that
Plasmodium strains are known to show great variability of their
proteins\cite{Mackinnon2010,Templeton2009}, thereby potentially
jeopardizing the efficiency of recognition by a highly specific
antibody developed for a rapid diagnostic test.

Beyond the scope of malaria research and diagnosis, our results can
contribute to biomedical applications of optically and/or
magnetically anisotropic submicron-sized particles
\cite{Koumura1999,Kuimova2009,Leonardo2012,Dharmadhikari2004,Dasgupta2011}.
We believe that the magnetic micro-rotor concept recognized
specifically for malaria pigment crystals can be generally applied
for the magnetic control, manipulation and detection of
submicron-sized magnetic particles functionalized to interact with
biomolecules and cells as it has already been demonstrated e.g. in
the study of the elastic properties of single DNA molecules using
magnetic beads \cite{Strick1996}.

Recently, various applications of rotating magnetic field has also
been proposed in material sciences. These include the three
dimensional alignment of magnetically anisotropic micro-particles
aiming to produce magnetically oriented microcrystal arrays for
X-ray crystallography \cite{F.Kimura_2010} or to reinforce
composites by superparamagnetic platelets \cite{R. M. Erb}. Rotating
magnetic fields have also been applied for investigating the
formation and rotational dynamics for chains of paramagnetic beads
\cite{A.K.Vuppu_2003,S.Melle_2000,S. Melle_2001}. Our method enables
the rotation of strong magnetic fields -- also characterized by a
high level of homogeneity over a large sample volume -- at
frequencies ranging from 0.1-130\,Hz, while the polarization
detection scheme supports monitoring of the dynamics even when
magnetic particles are present at ppm concentrations in solution.
Thus, this methodology can help to improve the performance in the
applications mentioned above.

\section*{Methods}
\textbf{Preparation and characterization.} Hemozoin crystals were
synthetized following the aqueous acid-catalyzed method described by
M. Jamarillo and co-workers \cite{Jamarillo2009}. Hemin was
dissolved in NaOH with the dropwise addition of propionic acid
adjusting a pH value of approximately 4. After annealing the mixture
for 18 hours at 70\,$^o$C, the crystals were separated and washed
with NaHCO$_{3}$, MilliQ water and MeOH, multiple times, alternately
-- as prescribed by the authors. According to our TEM measurements
the typical size of the crystals obtained by this method is
approximately 200$\times$200$\times$700\,nm$^3$.

The transmission electron micrographs were obtained by using two
different methods for the fixation of crystals. To avoid aggregation
of the crystals, in both cases the suspensions were prepared with
hemozoin contents $<$ 10\,$\mu$g/$\mu$l and long-term
ultrasonication. The aqueous suspension of hemozoin crystals was
dropped onto formvar membrane (purchased from Sigma-Aldrich) placed
on 200 mesh copper grid and dried at room temperature. The other
method applied for magnetically aligned ensembles of crystals was
freeze-fracture. In this case the suspension medium was a mixture of
70\% water and 30\% glycerol to prevent the formation of ice
crystals. The droplets (1-2\,$\mu$l) of suspension were pipetted on
a gold specimen holder kept in a field of B=0.5\,T at room
temperature for 30\,s, then plunged into partially solidified Freon
for 20\,s freezing and then placed and stored in liquid nitrogen.
Fracturing was carried out at 173\,K in a Balzers Freeze-fracture
Device (Balzers BAF 400 D). The fractured faces were etched for
30\,s at 173\,K. The replicas, prepared by platinum-carbon
shadowing, were cleaned and washed with distilled water. The
membranes and replicas obtained by the two methods were examined in
a transmission electron microscope as shown in
Fig.~\ref{fig:structure} and Fig.~S1, respectively.

For experiments performed on hemozoin crystals suspended in blood
plasma and hemolyzed blood, the blood plasma was obtained via the
centrifugation of blood samples and hemolysis of blood was achieved
by 20-fold dilution of blood with distilled water. Freshly drawn
blood was acquired from healthy volunteers.

\textbf{Magnetization measurements.} Magnetization measurements on
suspensions were performed using a Superconducting Quantum
Interference Device (SQUID) in magnetic fields ranging from B=0 to
5\,T at temperatures T=2-300\,K. Powder samples were measured over a
broader field range B=0-9\,T using a magnetometer with a.c. pick-up
coil. The magnetization component parallel to the applied magnetic
field was detected. Liquid samples were placed in hermetically
closed plastic straws, while gelatine capsules were used in the case
of the solid powder samples. The diamagnetic baselines originating
from the sample holders and the suspension medium were measured
separately and subtracted from the data.

\textbf{Magneto-optical methodology.} The spectrometer capable of
the measurement of MLD over the wavelength range of
$\lambda$=180-1300\,nm was assembled using a triple grating
monochromator, broad-band light sources (Xe-arc and tungsten lamps)
together with a photomultiplier and an InGaAs photodiode as
detectors. The experiment was set up in Voigt configuration, that is
the magnetic field was applied perpendicularly to the direction of
the light propagation. Light beam after the sample was collected
using lenses with typical numerical aperture of NA$=$0.1-0.2. The
fast switching between the light polarizations parallel and
perpendicular to the magnetic field was carried out with a fused
silica photoelastic modulator operating at a frequency of
50\,kHz\cite{Sato1981}. In order to eliminate linear polarization
effects other than MLD, the zero-field baseline was measured and
subtracted from the finite-field data.

\begin{figure}[h!]
\includegraphics[width=3.4in]{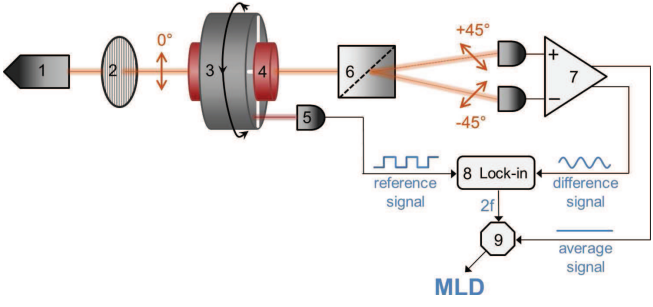}
\caption{\textbf{$\mid$ Flowchart of the diagnostic setup.} The beam
from the laser diode (1) passes through a polarizer (2) and becomes
vertically polarized. Then it goes through the sample holder (4)
located in the bore of the Halbach magnet (3). The magnet is rotated
with a frequency $f$ by a d.c. motor, thus, the uniform magnetic
field of B$\approx$1\,T at the sample position rotates within the
plane perpendicular to the light propagation. After the sample, the
beam is divided into two parts with orthogonal polarizations
($\pm$45$^o$) by a Rochon prism (6). The difference and the average
of their intensities are detected by a balanced photodiode bridge
(7). The $2f$ component of the difference signal is filtered out by
a lock-in amplifier (8) using the reference signal from an
optoswitch (5) monitoring the rotation of the magnet. To obtain the
MLD signal, the amplitude of the second harmonic ($2f$) signal is
normalized with the average signal by a divisor (9).}
\label{fig:sketch}
\end{figure}

\textbf{Diagnostic method and instrumentation.} The principles of
the new technique have been described in the main text.
Figure~\ref{fig:sketch} provides a schematic representation of the
diagnostic setup.

\textbf{Derivation of MLD effect for suspensions.} As discussed in
the main text, the magnetic and optical properties of hemozoin
crystals are characterized by an axial anisotropy. Hence, the value
of the complex transmission coefficient for light polarization
parallel to the C$_{4v}$ axis ($z$-axis) of a crystal differs from
the values corresponding to polarizations within the perpendicular
plane ($xy$-plane), i.e. $t_z$$\neq$$t_x$$=$$t_y$. The same
difference holds for the magnetization in external magnetic fields
pointing along the $z$-axis or lying in the $xy$-plane as
$M_z$$\neq$$M_x$$=$$M_y$.

While the matrix of the transmission coefficients, $\hat{t}$, is
diagonal for each crystal in its own $xyz$ frame, the transmission
of the whole suspension needs to be calculated in a common $x'y'z'$
reference frame, where $y'$- and $z'$-axis are conveniently chosen
as the direction of the light propagation and the magnetic field,
respectively. The new form of the transmission matrix, $\hat{t}'$ --
for a crystal with its $z$-axis pointing in the direction defined by
the azimuth angle, $\theta$ and polar angle, $\phi$ in the $x'y'z'$
frame -- can be obtained by the corresponding base transformation.
In this common frame, the portion of the light intensity transmitted
by the crystal for polarization along the $x'$/$z'$ direction is
given by $T_{x'/z'}=\left| \hat{t}' \cdot \bf{e}_{x'/z'}
\right|^{2}$, where $\bf{e}_{x'}$ and $\bf{e}_{z'}$ are unit vectors
pointing along the $x'$- and $z'$-axis, respectively.

For an ensemble of the crystals, the macroscopic linear dichroism
exhibited by the suspension is obtained by averaging over the
contributions from individual crystallites according to
$\frac{\Delta \mathcal{T}}{\mathcal{T}}=
c\cdot\frac{T_x-T_z}{T_x+T_z}\cdot\int_{0}^{\pi} \pi
f(\theta)(3\cos^{2}\theta-1)\sin{\theta}d\theta$, where
$T_x$$=$$|t_x|^2$ and $T_z$$=$$|t_z|^2$. The distribution of the
azimuth angle is governed by the Boltzmann factor:
$f(\theta)=\frac{e^{-U/k_BT}}{2\pi
\int_{0}^{\pi}e^{-U/k_BT}\sin{\theta}d\theta}$. The angular
distribution is independent of the polar angle, since the full
rotational symmetry around the magnetic field is preserved. Assuming
a linear field dependence of the magnetization, which is
experimentally confirmed at room temperature, the magnetic
anisotropy energy is
$U$$=$$-\frac{1}{2}\frac{B^2}{\mu_0}cos^2(\theta)(\chi_{zz}-\chi_{xx})V$.
Here, $\chi_{zz}$ and $\chi_{xx}$ stand for the linear magnetic
susceptibility of a crystal along the hard axis and within the easy
plane, respectively. The typical volume of a crystal, $V$, is
approximated as 200$\times$200$\times$700\,nm$^{3}$ according to TEM
images. The field dependence of $\frac{\Delta
\mathcal{T}}{\mathcal{T}}$ was evaluated numerically and used for
the fitting of the MLD data with two free parameters. The first
parameter is a scale factor, from which we could obtain the
transmission anisotropy of a single crystal,
$\frac{T_x-T_z}{T_x+T_z}$, as a function of the wavelength. The
other parameter is $(\chi_{zz}-\chi_{xx})V$, which determines the
distribution of the azimuth angle via the magnetic anisotropy
energy, hence, it is the only parameter describing the universal
field dependence of MLD. Considering an average crystal volume of
$V$$=$$2.8\cdot10^{-20}$\,m$^{3}$, the difference in the
magnetization densities is directly obtained.

\textbf{Numerical calculation of magnetization.} As argued in the
main text, the magnetic behaviour of $Fe^{3+}$ ions with S=5/2 spins
in hemozoin is described by the following axially symmetric
Hamiltonian: $H=
D\left(S_{z}^2-\frac{S(S+1)}{3}\right)+\mu_Bg\mathbf{BS}$. For a
given set of \{D, $\mathbf{B}$\}, we determined the energy
eigenvalues ($\varepsilon_{n}$) by the numerical diagonalization of
this 6$\times$6 matrix. Then, the magnetization density vector was
obtained according to
$\mathbf{M}=\frac{1}{V}k_{B}T\frac{\partial}{\partial
\mathbf{B}}lnZ$, where $Z$ is the partition function in the grand
canonical ensemble, $Z=\left(\sum_{n}{e^{-\varepsilon_{n}/k_BT}}
\right)^{N}$. Due to the magnetocrystalline anisotropy, the energy
eigenvalues depend not only on the strength of the magnetic field
but also on its orientation relative to the crystal ($\theta$). For
comparison with the experimental data, only the component of the
magnetization vector parallel to the field direction was considered.

Besides the principal values $M_x$ and $M_z$, the magnetization
density of unordered samples were also evaluated by averaging over
the contributions from individual crystals:
$\left<M\right>=\frac{1}{4\pi}\int_{0}^{\pi}2\pi[M_{x}(B_x)\sin{\theta}+M_{z}(B_z)\cos{\theta}]\sin{\theta}d\theta$.
$M_{x}$ and $\left<M\right>$ were respectively measured on oriented
and random hemozoin suspensions in a mixture of 70\% water and 30\%
glycerol. In the former case we used a field-cooled freezing
procedure. The experimental data, both the field and the temperature
dependent magnetization curves, were fitted using the single-ion
anisotropy factor (D) as the only fitting parameter.

The good correspondence between the values of axial anisotropy found
in the present magnetization experiments and reported by former
EPR\cite{Sienkiewicz2006} and M\"ossbauer\cite{Bohle1998}
spectroscopic studies implies that in paramagnetic hemozoin crystals
magnetocrystalline anisotropy dominates over the shape anisotropy
unlike in usual ferro- and ferrimagnetic crystals with micron or
submicron size. This is further supported by TEM images recorded
from cleaved surfaces of field-cooled suspensions as presented in
the Supplementary Information.

%
%

\textbf{Acknowledgements} We thank G. Mihaly, Y. Tokura, M.
Kellermayer and Sz. Osv\'ath for fruitful discussions. This work was
supported by Hungarian Research Funds OTKA PD75615, CNK80991,
CNK81056, Bolyai 00256/08/11, T\'AMOP-4.2.2.B- 10/1-2010-0009,
T\'AMOP-4.2.1.B-09/1/KMR-2010-0001, and ANR-NKTH ADD-MAL from
National Innovation Office.

\textbf{Author Contributions} A.B., A.O., L.K., A.B\'ota, I.K.
performed the measurements; A.B., A.O., S.B., I.K. analysed the
data; E. T-Sz. contributed to the sample preparation; A.B., A.O.,
V.K., D.Sz., I.K. developed the magneto-optical setup and the
diagnostic device; A.B., A.O., I.K. wrote the manuscript; each
author discussed the results; and I.K. planned and supervised the
project.

\textbf{Competing financial interests} The authors declare no
competing financial interests. \vspace{20mm}
\newpage


\begin{references}
%
\bibitem{WHO2011}World malaria report 2011, December 13, 2011.
(http://www.who.int/malaria).
%
\bibitem{Greenwood2005}Greenwood, B.M., Bojang, K., Whitty, C.J., Targett,
G.A. Malaria, {\it Lancet} {\bf 365}, 1487$-$1498 (2005).
%
\bibitem{Yuan2011}Yuan, J., Cheng, K.C., Johnson, R.L., Huang, R., Pattaradilokrat, S., Liu, A., Guha, R., Fidock, D.A., Inglese, J., Wellems, T.E., Austin, C.P. and
Su, X. Chemical genomic profiling for antimalarial therapies,
response signatures, and molecular targets, {\it Science} {\bf 333},
724$-$729 (2011).
%
\bibitem{Bell2006} Bell, D., Wongsrichanalai, C. and Barnwell,
J.W. Ensuring quality and access for malaria diagnosis: how can it
be achieved? {\it Nat. Rev. Microbiol.} {\bf 4}, 682$-$695 (2006).
%
\bibitem{Moody2002}Moody, A. Rapid diagnostic tests for malaria
parasites, {\it Clin. Microbiol. Rev.} {\bf 15} 66$-$78 (2002).
%
\bibitem{Wilson2012}Wilson, M.L. Malaria rapid diagnostic tests, {\it
Clin. Inf. Dis.} {\bf 54}, 1637$-$1641 (2012).
%
\bibitem{Murray2009}Murray, C.K. and Bennett, J.W. Rapid Diagnosis of Malaria, {\it
Interdiscip. Perspect. Infect. Dis.} {\bf 2009}, 415953 (2009).
%
\bibitem{Gamboa2010}Gamboa, D., Ho, M-F., Bendezu, J., Torres, K., Chiodini, P.L., Barnwell, J.W., Incardona, S., Perkins, M., Bell, D., McCarthy, J. and Cheng, Q. A Large Proportion of P. falciparum Isolates in the Amazon Region of Peru Lack pfhrp2 and pfhrp3: Implications for Malaria Rapid Diagnostic Tests
{\it PLoS ONE} {\bf 5}, e8091 (2010).
%
\bibitem{Rock1987}Rock, E.P., Marsh, K., Saul, A.J., Wellems, T.E., Taylor, D.W., Maloy,
W.L. and Howard, R.J. Comparative analysis of the Plasmodium
falciparum histidine-rich proteins HRP-I, HRP-II and HRP-III in
malaria parasites of diverse origin, {\it Parasitology} {\bf 95},
209$-$227 (1987).
%
\bibitem{Andrade2010}Andrade, B.B., Reis-Filho, A., Barros, A.M., Souza-Neto, S.M., Nogueira, L.L.,
Fukutani, K.F., Camargo, E.P., Camargo, L.M., Barral, A., Duarte, A.
and Barral-Netto, M. Towards a precise test for malaria diagnosis in
the Brazilian Amazon: comparison among field microscopy, a rapid
diagnostic test, nested PCR, and a computational expert system based
on artificial neural networks, {\it Malaria J.} {\bf 9}, 117 (2010).
%
\bibitem{Khairnar2009}Khairnar, K., Martin, D., Lau, R., Ralevski, F. and Pillai D.R.
Multiplex real-time quantitative PCR, microscopy and rapid
diagnostic immuno-chromatographic tests for the detection of
Plasmodium spp: performance, limit of detection analysis and quality
assurance, {\it Malaria J.} {\bf 8}, 284 (2009).
%
\bibitem{Zimmerman2006}Zimmerman, P.A., Thomson, J.M., Fujioka, H., Collins, W.E. and
Zborowski, M. Diagnosis of malaria by magnetic deposition
microscopy, {\it Am. J. Trop. Med. Hyg.} {\bf 74}, 568$-$572 (2006).
%
\bibitem{Belisle2008}B\'elisle, J.M., Costantino, S., Leimanis, M.L., Bellemare, M.-J.,
Bohle, D.S., Georges, E. and Wiseman, P.W. Sensitive detection of
malaria infection by third harmonic generation imaging, {\it
Biophys. J.} {\bf 94}, L26$-$L28 (2008).
%
\bibitem{Yatsushiro2010}Yatsushiro, S., Yamamura, S., Yamaguchi, Y., Shinohara, Y.,
Tamiya, E., Horii, T., Baba, Y. and Kataoka, M. Rapid and highly
sensitive detection of malaria-infected erythrocytes using a cell
microarray chip, {\it PloS ONE} {\bf 5}, e13179 (2010).
%
\bibitem{Samson2012}Samson, E.B., Goldschmidt, B.S., Whiteside, P.J.D., Sudduth,
A.S.M., Custer, J.R., Beerntsen, B. and Viator, J.A. Photoacoustic
spectroscopy of $\beta$-hematin, {\it J. Opt.} {\bf 14}, 065302
(2012).
%
\bibitem{Kim2010}Kim, C.C., Wilson, E.B. and DeRisi, J.L. Improved methods for magnetic purification of
malaria parasites and haemozoin, {\it Malaria J.} {\bf 9}, 17
(2010).
%
\bibitem{Newman2008}Newman, D.M., Heptinstall, J., Matelon, R.J., Savage, L., Wears, M.L., Beddow,
J., Cox, M., Schallig, H.D.F.H. and Mensz P.F. A magneto-optic route
toward the in vivo diagnosis of malaria: preliminary results and
preclinical trial data, {\it Biophys. J.} {\bf 95}, 994$-$1000
(2008).
%
\bibitem{Mens2010}Mens, P.F., Matelon, R.J., Nour, B.Y.M., Newman, D.M. and Schallig, H.D.F.H. Laboratory evaluation on the sensitivity and
specificity of a novel and rapid detection method for malaria
diagnosis based on magneto-optical technology (MOT), {\it Malaria
J.} {\bf 9}, 207 (2010).
%
\bibitem{Castilho2011}Castilho, M.S., Laube, T., Yamanaka, H., Alegret, S. and
Pividori, M.I. Magneto immunoassays for Plasmodium falciparum
histidine-rich protein 2 related to malaria based on magnetic
nanoparticles, {\it Anal. Chem.} {\bf 83}, 5570$-$5577 (2011).
%
\bibitem{Yuen2012}Yuen, C., Liu, Q. Magnetic field enriched surface enhanced resonance Raman spectroscopy for early malaria
diagnosis, {\it J. Biomed. Opt.} {\bf 17}, 017005 (2012).
%
\bibitem{Goldberg1990}Goldberg, D.E., Slater, A.F.G., Cerami, A. and Henderson, G.B. Hemoglobin degradation in the malaria parasite Plasmodium
falciparum: An ordered process in a unique organelle, {\it Proc.
Natl. Acad. Sci. USA} {\bf 87}, 2931$-$2935 (1990).
%
\bibitem{Gluzmann1994}Gluzman, I.V., Francis,  S.E., Oksman, A., Smith, C.E., Duffin, K.L and Goldberg,
D.E. Order and specificity of the Plasmodium falciparum hemoglobin
degradation pathway, {\it J. Clin. Invest.} {\bf 93}, 1602$-$1608
(1994).
%
\bibitem{Slater1991}Slater, A.F.G., Swiggard, W.J., Orton, B.R.,
Flitter, W.D., Goldberg, D.E., Cerami, A. and Henderson, G.B. An
iron-carboxylate bond links the heme units of malaria pigment, {\it
Proc. Natl. Acad. Sci. USA} {\bf 88}, 325$-$329 (1991).
%
\bibitem{Bohle1998}Bohle, D.S., Debrunner, P., Jordan, P.A., Madsen, S.K. and
Schulz C.E. Aggregated heme detoxification byproducts in malarial
trophozoites: $\beta-$hematin and malaria pigment have a single
S=5/2 iron environment in the bulk phase as determined by EPR and
magnetic M\"ossbauer spectroscopy, {\it J. Am. Chem. Soc.} {\bf
120}, 8255-8256 (1998).
%
\bibitem{Sienkiewicz2006}Sienkiewicz, A., Krzystek, J., Vileno, B., Chatain, G.,
Kosar, A.J., Bohle, D.S. and Forr\'o, L. Multi-frequency high-field
EPR study of iron centers in malarial pigments, {\it J. Am. Chem.
Soc.} {\bf 128}, 4534$-$4535 (2006).
%
\bibitem{Walczak2005}Walczak, M., Lawniczak-Jablonska, K., Sienkiewicz, A., Demchenko,
I.N., Piskorska, E., Chatain, G. and Bohle, D.S. Local environment
of iron in malarial pigment and its substitute $\beta-$hematin, {\it
Nucl. Instrum. Meth. B} {\bf 238}, 32$-$38 (2005).
%
\bibitem{Hanscheid2007}Hanscheid, T., Egan, T.J. and Grobusch, M.P. Haemozoin: from
melatonin pigment to drug target, diagnsotic tool, and immune
modulator. {\it Lancet Infect. Dis.} {\bf 7}, 675$-$685 (2007).
%
\bibitem{Jamarillo2009}Jaramillo, M., Bellemare, M.J., Martel, C., Shio, M.T.,
Contreras, A.P., Godbout, M., Roger, M., Gaudreault, E., Gosselin,
J., Bohle, D.S. and Olivier, M. Synthetic plasmodium-Like hemozoin
activates the immune response: A morphology - function study, {\it
PLoS ONE} {\bf 4}, e6957 (2009).
%
\bibitem{Pagola2000}Pagola, S., Stephens, P.W., Bohle, D.S., Kosar, A.D. and
Madsen, S.K. The structure of malaria pigment $\beta$-haematin, {\it
Nature} {\bf 404}, 307$-$310 (2000).
%
\bibitem{Noland2003}Noland, G.S., Briones, N. and Sullivan Jr., D.J.
The shape and size of hemozoin crystals distinguishes diverse
Plasmodium species, {\it Mol. Biochem. Parasit.} {\bf 130}, 91$–$99
(2003).
%
\bibitem{Oliveiraa2005}Oliveiraa, M.F., Kyciab, S.W., Gomezb, A., Kosarc, A.J.,
Bohlec, D.S., Hempelmannd, E., Menezese, D., Vannier-Santose, M.A.,
Oliveiraa, P.L. and Ferreira, S.T. Structural and morphological
characterization of hemozoin produced by Schistosoma mansoni and
Rhodnius prolixus, {\it FEBS Lett.} {\bf 579}, 610$-$616 (2005).
%
\bibitem{Bohle1993}Bohle, D.S. and Helms, J.B. Synthesis of
$\beta-$hematin by dehydrohalogenation of hemin, {\it Biochem.
Bioph. Res. Co.} {\bf 193}, 504$-$508 (2003).
%
\bibitem{Frosch2007}Frosch, T., Koncarevic, S., Zedler, L., Schmitt, M.,
Schenzel, K., Becker, K. and Popp J. In situ localization and
structural analysis of the malaria pigment hemozoin, {\it J. Phys.
Chem. B} {\bf 111}, 11047$-$11056 (2007).
%
\bibitem{Buller2002}Buller, R., Peterson, M.L., Almarsson, \~O. and
Leiserowitz, L. Quinoline binding site on malaria pigment crystal: a
rational pathway for antimalaria drug design, {\it Cryst. Growth.
Des.} {\bf 2}, 553$–$562 (2002).
%
\bibitem{Kapishnikov2012}Kapishnikov, S., Berthing, T., Hviid, L.,
Dierolf, M., Menzel, A., Pfeiffer, F., Als-Nielsen, J. and
Leiserowitz L. Aligned hemozoin crystals in curved clusters in
malarial red blood cells revealed by nanoprobe X-ray Fe fluorescence
and diffraction, {\it Proc. Natl. Acad. Sci. USA} {\bf 109},
11184$–$11187 (2012).
%
\bibitem{Eaton1978}Eaton, W.A., Hanson, L.K., Stephens, P.J., Sutherland,
J.C. and Dunn, J.B.R. J. Optical spectra of oxy- and deoxyhemoglobin
{\it J. Am. Chem. Soc.} {\bf 100}, 4991$-$5003 (1978).
%
\bibitem{Wood2004}Wood, B.R., Langford, S.J., Cooke, B.M., Lim, J., Glenister,
F.K., Duriska, M., Unthank, J.K. and McNaughton, D. Resonance Raman
spectroscopy reveals new insight into the electronic structure of
$\beta$-hematin and malaria pigment, {\it J. Am. Chem. Soc.} {\bf
126}, 9233$-$9239 (2004).
%
\bibitem{Newman2008_2}Newman, D.M., Matelon, R.J., Wears, M.L., Savage, L., Heptinstall, J., Beddow, J. and Cox,
M. Magneto-Optics in the Service of Medicine - Diagnosis via the
Cotton-Mouton Effect-, {\it PhotonicsGlobal@Singapore, 2008. IPGC
2008. IEEE}, 1$-$3 (2008).
%
\bibitem{Langevin1910}Langevin, P. Magnetisme et theorie des electrons, {\it CR Acad. Sci. Paris} {\bf
151}, 331$–$368 (1910).
%
\bibitem{Butykai}Butykai, A. {\it et al.} to be published.
%
\bibitem{Halbach1980}Halbach, K. Design of permanent multipole magnets with oriented rare earth cobalt
material, {\it Nucl. Instrum. Methods} {\bf 169}, 1$-$10 (1980).
%
\bibitem{Erb_2012_Soft}Erb, R.M., Segmehl, J., Charilaou, M., Leoffler J.F. and Studart A.R.
Non-linear alignment dynamics in suspensions of platelets under
rotating magnetic fields, {\it Soft Matter}, \textbf{8}, 7604–7609,
(2012).
%
\bibitem{Tierno_2009}Tierno, P., Claret, J. and Sagués, F. Overdamped dynamics of paramagnetic ellipsoids in a precessing magnetic field, {\it Phys. Rev. E} \textbf{79}, 021501
(2009).
%
\bibitem{Dhar_2007}Dhar, P., Swayne, C.D., Fischer, T.M., Kline, T. and Sen A. Orientations of over damped magnetic nano-gyroscopes, {\it Nano Lett.} \textbf{7}, 1010-1012
(2007).
%
\bibitem{R. M. Erb}Erb, R.M., Libanori, R., Rothfuchs N. and Studart, A.R. Fields
Composites Reinforced in Three Dimensions by Using Low Magnetic
Fields, {\it Science} \textbf{335}, 199 (2012).
%
\bibitem{Mackinnon2010}Mackinnon, M.J. and Marsh, K. The selection landscape of malaria
parasites, {\it Science} {\bf 328}, 866-871 (2010).
%
\bibitem{Templeton2009}Templeton, T.J. The varieties of gene amplification,
diversification and hypervariability in the human malaria parasite,
Plasmodium falciparum, {\it Mol. Biochem. Parasit.} {\bf 166},
109-116 (2009).
%
\bibitem{Koumura1999}Koumura, N., Zijlstra, R.W.J., van Delden, R.A., Harada, N. and
Feringa, B.L. Light-driven monodirectional molecular rotor, {\it
Nature} {\bf 401}, 152-155 (1999).
%
\bibitem{Kuimova2009}Kuimova, M.K., Botchway, S.W., Parker, A.W., Balaz, M., Collins,
H.A., Anderson, H.L., Suhling, K. and Ogilby, R. Imaging
intracellular viscosity of a single cell during photoinduced cell
death, {\it Nat. Chem.} {\bf 1} 69$-$73 (2009).
%
\bibitem{Leonardo2012}Di Leonardo, R., B\'uz\'as, A., Kelemen, L., Vizsnyiczai, G., Oroszi, L. and
Ormos, P. Hydrodynamic Synchronization of Light Driven Microrotors,
{\it Phys. Rev. Lett.} {\bf 109}, 034104 (2012).
%
\bibitem{Dharmadhikari2004}Dharmadhikari, J.A., Roy, S., Dharmadhikari, A.K., Sharma, S. and
Mathur, D. Naturally occurring, optically driven, cellular rotor
{\it Appl. Phys. Lett.} {\bf 85}, 6048$-$6050 (2004).
%
\bibitem{Dasgupta2011}Dasgupta, R., Ahlawat, S., Verma, R.S. and Gupta P.K. Optical
orientation and rotation of trapped red blood cells with
Laguerre-Gaussian mode, {\it Opt. Express} {\bf 19}, 7680$-$7688
(2011).
%
\bibitem{Strick1996}Strick, T.R., Allemand, J.-F., Bensimon, D., Bensimon, A. and
Croquette, V. The Elasticity of a Single Supercoiled DNA Molecule,
{\it Science} {\bf 271}, 1835$-$1837 (1996).
%
\bibitem{F.Kimura_2010}Kimura, F., Mizutani, K., Mikami, B. and Kimura T. Single-Crystal X-ray Diffraction
Study of a Magnetically Oriented Microcrystal Array of Lysozyme,
{\it Cryst. Growth Des.} {\bf 11}, 12-15 (2011).
%
\bibitem{A.K.Vuppu_2003}Vuppu, A.K., Garcia, A.A. and Hayes, M.A. Video Microscopy of Dynamically Aggregated
Paramagnetic Particle Chains in an Applied Rotating Magnetic Field,
{\it Langmuir}, {\bf 19}, 8646-8653 (2003).
%
\bibitem{S.Melle_2000}Melle, S., Fuller, G.G. and Rubio, M.A. Structure and dynamics of magnetorheological fluids in rotating magnetic fields, {\it Phys. Rev. E} {\bf 61}, 4111-4117
(2000).
%
\bibitem{S. Melle_2001}Melle, S., Rubio, M.A., Fuller, G.G. Orientation dynamics of magnetorheological fluids subject to rotating external fields, {\it Int. J. Mod. Phys. B} {\bf 15}, 758-766
(2001).
%
\bibitem{Sato1981}Sato, K. Measurement of magneto-optical Kerr effect using piezo-birefringent
modulator, {\it Jpn. J. Appl. Phys.} {\textbf 20}, 2403$-$2409
(1981).\\ \vspace{10mm}
%
\end{references}
\end{document}